# Observations of Slow Solar Wind from Equatorial Coronal Holes

Y.-M. Wang and Y.-K. Ko
Space Science Division, Naval Research Laboratory, Washington, DC 20375, USA; yi.wang@nrl.navy.mil, yuan-kuen.ko@nrl.navy.mil
Received 2019 May 6; revised 2019 June 3; accepted 2019 June 17; published 2019 August 5

## Abstract

Because of its distinctive compositional properties and variability, low-speed ($\lesssim 450\ \mathrm{km\ s^{-1}}$) solar wind is widely believed to originate from coronal streamers, unlike high-speed wind, which comes from coronal holes. An alternative scenario is that the bulk of the slow wind (excluding that in the immediate vicinity of the heliospheric current sheet) originates from rapidly diverging flux tubes rooted inside small coronal holes or just within the boundaries of large holes. This viewpoint is based largely on photospheric field extrapolations, which are subject to considerable uncertainties and do not include dynamical effects, making it difficult to be certain whether a source is located just inside or outside a hole boundary, or whether a high-latitude hole will be connected to Earth. To minimize the dependence on field-line extrapolations, we have searched for cases where equatorial coronal holes at central meridian are followed by low-speed streams at Earth. We describe 14 examples from the period 2014–2017, involving Fe XIV 21.1 nm coronal holes located near active regions and having equatorial widths of ~3°–10°. The associated in situ wind was characterized by speeds $v \sim 300$–$450\ \mathrm{km\ s^{-1}}$ and by $\mathrm{O^{7+}/O^{6+}}$ ratios of ~0.05–0.15, with $v$ showing the usual correlation with proton temperature. In addition, consistent with other recent studies, this slow wind had remarkably high Alfvénicity, similar to that in high-speed streams. We conclude that small coronal holes are a major contributor to the slow solar wind during the maximum and early post-maximum phases of the solar cycle.

## 1. Introduction

Compared to fast solar wind, slow wind, defined here as having near-Earth speeds $v \lesssim 450\ \mathrm{km\ s^{-1}}$, is characterized by greater variability, lower and more isotropic proton temperatures, higher ion freeze-in temperatures, greater enrichment in elements of low first-ionization potential (FIP), smaller He/H ratios near sunspot minimum, and a tendency toward lower Alfvénicity (see the review of Abbo et al. 2016). Such differences have led to the standard bimodal picture in which fast wind originates from coronal holes, whereas slow wind escapes from coronal streamers.

Direct evidence for the streamer origin of at least one component of the slow wind comes from coronagraph observations of small flux ropes ("blobs") that are continually emitted from the cusps of helmet streamers and travel outward along the heliospheric current sheet (HCS) and its surrounding plasma sheet (see, e.g., Sheeley et al. 2009). However, low-speed wind is often observed at large angular distances from the HCS, especially around sunspot maximum. According to the S-web model of Antiochos et al. (2011), a major contribution to the slow wind comes from the boundaries between like-polarity coronal holes, where "pseudostreamers" undergo interchange reconnection with the adjacent open flux. In this scenario, all of the slow wind is associated with the network of separatrices dividing open flux domains of the same or opposite polarity.

In an alternative scenario, the bulk of the low-speed wind comes from rapidly diverging flux tubes rooted inside small coronal holes or just within the boundaries of large holes. This view is based on the inverse correlation found between the wind speed at 1 au and the rate at which the underlying coronal flux tube expands, as inferred from outward extrapolations of photospheric field measurements (see, e.g., Wang & Sheeley 1990; Arge & Pizzo 2000; Poduval 2016). The relationship may be understood physically if the energy deposition rate is assumed to depend on the local coronal field strength. The effect of rapid flux-tube divergence is then to concentrate the heating near the coronal base, increasing the mass flux but decreasing the energy per proton available to accelerate the wind. The ion charge-state ratios likewise depend on the location and strength of the heating along the open flux tubes (Wang & Sheeley 2003; Cranmer et al. 2007; Wang et al. 2009; Oran et al. 2015). Recent support for this scenario comes from an analysis of 1974–1978 *Helios* data by Stansby et al. (2019), who found that most of the low-speed (~200–500 km s⁻¹) wind at 0.3–0.4 au has high Alfvénicity, interpreted as a signature of outflow along open field lines (see also Marsch et al. 1981; Roberts et al. 1987a). In addition, D'Amicis & Bruno (2015) and D'Amicis et al. (2019) found evidence for Alfvénic slow wind at 1 au during the cycle 23 maximum, which they associated with coronal hole boundaries or small coronal holes.

The inference that coronal holes provide a major contribution to the slow solar wind depends largely on extrapolations of the photospheric magnetic field, which are subject to considerable uncertainties and do not include effects such as field-line reconnection. In the widely used potential-field source-surface (PFSS) model, all of the flux beyond the source surface is by definition "open," so that Earth is always connected to an open field region (representing a coronal hole) unless it is located at the source-surface neutral line (which corresponds to the HCS). Because the slow wind originates near coronal hole boundaries in both of the competing scenarios, field-line tracing may not be sufficiently accurate to distinguish between them, even with the use of more sophisticated methods (such as magnetohydrodynamic simulations) for extrapolating the photospheric field.

Near solar minimum, the PFSS model predicts that much of the slow wind at Earth comes from just inside the polar coronal holes, with the open field lines bending from high latitudes into the ecliptic. In contrast, the prevailing view is that this wind originates from inside helmet streamers, which indeed lie close to the ecliptic plane at this time, as does their heliospheric extension in the form of the HCS. According to Zhao et al. (2009), the value of the oxygen charge-state ratio that separates coronal hole from non-coronal hole wind is $O^{7+}/O^{6+} \simeq 0.145$. However, during the 2007–2009 activity minimum, the median value of $O^{7+}/O^{6+}$ recorded by the Solar Wind Ion Spectrometer (SWICS) on the *Advanced Composition Explorer* (*ACE*) fell to only ∼0.06, with the decrease being correlated with that of the underlying photospheric field strength (see Zhao et al. 2014; Wang 2016). This suggests that compositional properties alone are not sufficient to determine the sources of the solar wind (see also Stakhiv et al. 2015).

To help resolve the question of whether coronal holes are a major contributor to the slow wind, it is important to minimize the dependence on field-line tracing when locating the source of an observed wind stream. In general, we may assume that coronal holes lying close to the ecliptic or solar equator are likely to become connected to Earth a few days after their central-meridian passage. Moreover, the wind from the centers of these holes is more likely to be sampled than in the case of high-latitude holes. Because they typically form within the unipolar remnants of low-latitude active regions (ARs), equatorial holes are often observed during the maximum and early declining phases of the sunspot cycle.

Employing extreme-ultraviolet (EUV) images from the *Solar Dynamics Observatory* (*SDO*) and near-Earth measurements of the solar wind, we have identified the wind streams associated with equatorial coronal holes observed during 2014–2017. This paper describes cases in which the peak speeds are less than or on the order of $450\,\mathrm{km\,s^{-1}}$. As predicted by earlier studies, the open flux rooted in these holes tends to diverge more rapidly than in the case of equatorial holes that produce fast wind. Moreover, the associated low-speed wind is characterized by high levels of Alfvénic fluctuations and by relatively low charge-state ratios. Our results are interpreted as providing strong support for the assertion that coronal holes are sources of both slow and fast solar wind.

## 2. Procedure

To identify coronal holes, we use observations recorded in the Fe XIV 21.1 nm emission line by the Atmospheric Imaging Assembly (*SDO*/AIA). To expedite the search for equatorial holes during 2014–2017, central-meridian strips are extracted from full-disk images and assembled into 27.3 day Carrington maps.

For comparison with the observed distribution of dark EUV coronal holes, we construct corresponding maps of open field regions by applying PFSS extrapolations to photospheric field measurements from the Wilcox Solar Observatory (WSO) or the Global Oscillation Network Group (NSO/GONG2). Such extrapolations do not take into account dynamical processes (such as interchange reconnection) that may occur at the interfaces between open and closed flux; however, the results are generally in agreement with steady-state magnetohydrodynamical computations (see, e.g., Neugebauer et al. 1998; Riley et al. 2006; Cohen 2015). In our version of the PFSS model, the magnetic field $\boldsymbol{B}(r, L, \phi)$ remains current-free from the coronal base to $r = R_{ss} = 2.5\,R_\odot$, where the transverse field components are required to vanish; here $r$ denotes heliocentric distance, $L$ heliographic latitude, and $\phi$ Carrington longitude. At $r = R_\odot$, $B_r$ is matched to the photospheric field, assumed to be approximately radial at the depth where it is measured. The WSO fluxes are corrected for the saturation of the Fe I 525.0 nm line profile by multiplying them by the latitude-dependent factor $(4.5–2.5\sin^2 L)$ (Wang & Sheeley 1995). The GONG2 fields are scaled upward by 1.65 so that their total dipole strengths are consistent with NSO/SOLIS measurements.

The factor by which a flux tube expands in (Sun-centered) solid angle between the coronal base and radius $r$ is given by

$$f(r) = \left(\frac{R_\odot}{r}\right)^2 \frac{B_0}{|B_r|}, \qquad (1)$$

where $B_0$ denotes the footpoint field strength. To characterize the divergence rates of open field lines, we employ the parameter $f_{max}$, defined as the maximum value of $f(r)$ between $r = R_\odot$ and $r = R_{ss}$.

For comparisons with in situ data, we assume that the solar wind propagates radially from $r = 2.5\,R_\odot$ to 1 au, with the transit time being inversely proportional to the smoothed near-Earth proton speed $v$. Hourly averages of $v$, proton density $n_p$, proton temperature $T_p$, and interplanetary magnetic field (IMF) longitude angle $\phi_B$ are extracted from the OMNIWeb site,[1] while two-hourly averages of $O^{7+}/O^{6+}$ and Fe/O (where Fe represents a low-FIP element) are obtained from the *ACE* SWICS 2.0 database.[2] It should be noted that the SWICS 2.0 measurements of $O^{7+}/O^{6+}$ have an instrumental floor of ∼0.05 and that values ≲0.1 are uncertain by factors of order 2. Also, the SWICS 2.0 values of Fe/O are typically a factor of 2 higher than the pre-2011 SWICS 1.1 values.

Employing 64 s averages of $\boldsymbol{v}$ and $\boldsymbol{B}$ (in *RTN* coordinates) from the *ACE* MAG/SWEPAM database, we also calculate the Alfvénicity proxy $C_{vB} \equiv (\delta\boldsymbol{v} \cdot \delta\boldsymbol{B})/(|\delta v||\delta B|)$, which measures the degree of correlation between velocity and magnetic fluctuations (see Roberts et al. 1987b). Here we take $\delta v = v - v_{12}$ and $\delta B = B - B_{12}$, where the subscript "12" signifies a 12 minute running mean. The variation of $C_{vB}$ is very similar to that of the cross-helicity $\sigma_c$, but does not include the contribution of the proton density, which is subject to frequent dropouts in the SWEPAM database (see Ko et al. 2018).

For consistency with the spatial/temporal resolution of our PFSS extrapolations, the in situ parameters are smoothed by taking 10 hr running averages. The source locations of the near-Earth wind are found by tracing inward along the magnetic field, taking into account the longitude shift due to solar rotation.

## 3. Slow Wind from Equatorial Coronal Holes During 2014–2017

We have searched for coronal holes that satisfy the following criteria: (1) the associated Fe XIV 21.1 nm structure straddles the solar equator and/or the ecliptic plane; (2) wind speeds ≲450 km s$^{-1}$ are recorded near Earth after the hole crosses the disk center (with the lag being roughly consistent with the Sun–Earth transit time); (3) the PFSS extrapolation shows a corresponding open field region that is connected to the ecliptic; and (4) no obvious coronal mass ejection (CME) signatures (such as values of $O^{7+}/O^{6+} \gtrsim 0.5$) are present in the in situ data.

---

## WSO 2147 (2014 FEBRUARY 11 - MARCH 11)

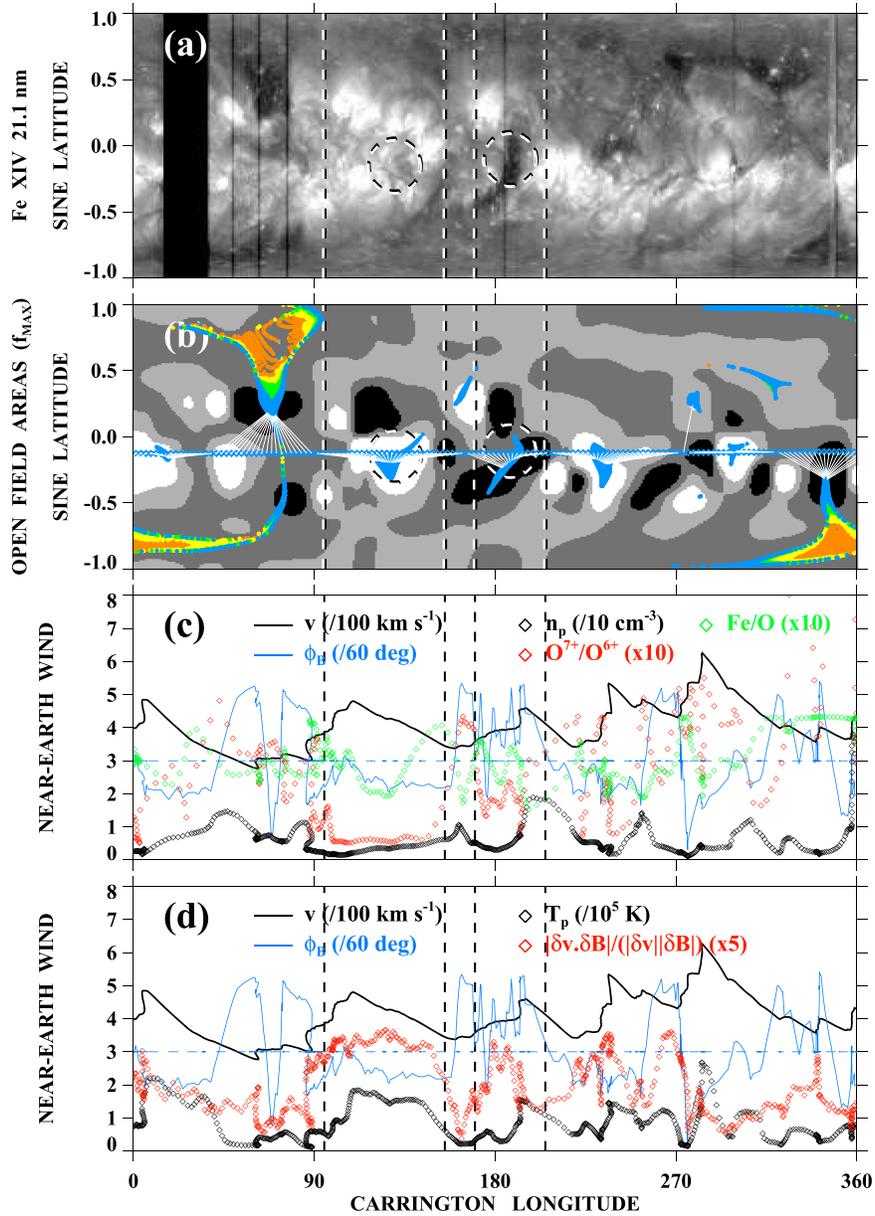

**Figure 1.** Two equatorial coronal holes (circled) that were sources of slow wind during CR 2147. (a) Distribution of Fe XIV 21.1 nm intensity as recorded by *SDO*/AIA. The equatorial hole at longitude $\phi \sim 130°$ is hidden beneath the background AR loops (see Figure 2). (b) WSO photospheric field (saturated at $B_r = \pm 10$ G) with PFSS-derived open field regions overplotted as colored dots. Colored diamonds (plotted in the ecliptic plane) indicate the source surface longitudes of Earth-directed flux tubes, with white lines connecting them to their photospheric footpoints. Dots and diamonds are color-coded as follows: blue ($f_{max} > 18$); green ($9 < f_{max} < 18$); yellow ($4 < f_{max} < 9$); orange ($f_{max} < 4$). (c) Variation of the near-Earth wind speed $v$, IMF longitude angle $\phi_B$, proton density $n_p$, $O^{7+}/O^{6+}$, and Fe/O as a function of Carrington longitude $\phi$ (time runs from right to left). The $O^{7+}/O^{6+}$ measurements have an instrumental floor value of $\sim 0.05$. Horizontal dashed line marks $\phi_B = 180°$; $\phi_B > 180°$ ($\phi_B < 180°$) corresponds to inward (outward) IMF. (d) Variation of $v$, $\phi_B$, proton temperature $T_p$, and the Alfvénicity proxy $|C_{vB}| \equiv |\delta\mathbf{v} \cdot \delta\mathbf{B}|/(|\delta v||\delta B|)$ as a function of $\phi$. The OMNI and *ACE* data have been smoothed by taking 10 hr running means and mapped back to the source surface assuming a transit time $\propto v^{-1}$. The wind speeds from the equatorial holes range from $\sim 350$ to $\sim 480$ km s$^{-1}$, consistent with the large expansion factors of the Earth-directed flux tubes (blue diamonds in (b)). Despite the low speeds, the Alfvénicities reach levels as high as $|C_{vB}| \sim 0.6$–$0.7$.

### 3.1. Carrington Rotation (CR) 2147: 2014 February–March

We begin by describing two equatorial sources of slow wind during CR 2147, near the maximum of solar cycle 24. Figure 1 displays (a) the global distribution of Fe XIV 21.1 nm emission, (b) the WSO photospheric field with the footpoint areas of open flux represented by colored dots, (c) the variation of $v$, $\phi_B$, $n_p$, $O^{7+}/O^{6+}$, and Fe/O as a function of Carrington longitude $\phi$, and (d) the variation of $v$, $\phi_B$, $T_p$, and $|C_{vB}| \equiv |\delta\mathbf{v} \cdot \delta\mathbf{B}|/(|\delta v||\delta B|)$ plotted against $\phi$. In the map of open field regions, the darker colors (blue, green) represent larger values of $f_{max}$, while the warmer colors (yellow, orange) denote smaller values; the colored diamonds (similarly coded) indicate the maximum expansion factors of flux tubes connected to the ecliptic plane, with white lines linking their ecliptic positions to their photospheric footpoints. Note that the solar wind data in the bottom two panels are plotted in time-reversed order, and that the wind has been mapped back to the source surface assuming a time lag proportional to $v^{-1}$. The pairs of vertical dashed lines in Figure 1 and subsequent figures indicate the approximate longitudinal extents of the wind streams associated with the circled equatorial holes; the actual stream boundaries are often difficult to determine and may differ from the marked positions by as much as ~5°–10°.

The EUV map in Figure 1(a) shows a dark coronal hole that intersects the equator/ecliptic near $\phi \sim 190°$ and has a longitudinal width of $\Delta\phi \sim 6°$. The corresponding open field region in Figure 1(b) is embedded in a negative unipolar area and is connected to the ecliptic plane. The in situ measurements (Figure 1(c)) show a small wind stream near $\phi \sim 190°$, whose speeds are in the range ~360–460 km s$^{-1}$ and which has the expected inward IMF polarity ($\Phi_B > 180°$). The O$^{7+}$/O$^{6+}$ ratio (red diamonds) averages around ~0.15–0.2 inside the stream and increases steeply toward the sector boundaries, while the values of Fe/O (green diamonds) fluctuate between ~0.15 and ~0.35. The large variations in O$^{7+}$/O$^{6+}$ and Fe/O at longitudes $\gtrsim 210°$ reflect the presence of AR-associated transient activity to the west of the coronal hole.

At longitudes ~95°–155°, Figure 1(c) shows a clearly defined stream with $v \sim 350$–480 km s$^{-1}$ and outward IMF polarity ($\Phi_B < 180°$). The O$^{7+}$/O$^{6+}$ ratio remains close to the SWICS 2.0 floor value of ~0.05 across the main body of the stream, but rises steeply near its edges; the values of Fe/O vary from ~0.2 near the middle of the stream to ~0.4 toward its edges. A corresponding open field region, embedded in a positive unipolar area and connected to the ecliptic plane, appears in Figure 1(b). However, no counterpart coronal hole seems to be present in Figure 1(a), where only a slight darkening of the Fe XIV emission is seen within the circled area. To clarify the nature of this structure, we tracked it in individual 21.1 nm images as it rotated toward central meridian during 2014 February 26–28. When located to the east of the disk center, and thus viewed at an angle (Figures 2(a) and (b)), the feature appears sufficiently dark to be recognizable as a small coronal hole; the contrast with the background emission decreases near disk center (Figure 2(c)). This behavior may be attributed to the fanning-out of the adjacent AR loops, which become oriented transverse to the line of sight when the hole is at central meridian (for other examples, see Wang 2017).

As indicated by the circled areas in Figure 1(b), the open field regions corresponding to the two equatorial holes are both characterized by large expansion factors ($f_{max} > 18$: blue pixels), consistent with the low speeds observed at Earth.

From Figure 1(d), we see that the wind streams associated with both equatorial holes have high Alfvénicities $|C_{vB}| \sim 0.6$–0.7, despite their relatively low speeds. This result is consistent with the findings of D'Amicis et al. (2019), who hypothesized (without explicitly identifying the individual sources) that the episodes of slow Alfvénic wind that they detected during the 2000–2002 activity maximum originated from small, low-latitude coronal holes.

### 3.2. CR 2148: 2014 March–April

Figure 3 displays the distribution of Fe XIV intensity during CR 2148, the WSO photospheric field with the footpoint areas of open flux overplotted, and the variation of $v$, $\phi_B$, $n_p$, O$^{7+}$/O$^{6+}$, Fe/O, $T_p$, and $|C_{vB}| \equiv |\delta v \cdot \delta B|/(|\delta v||\delta B|)$ as a function of Carrington longitude. Throughout this interval, the in-ecliptic wind is dominated by speeds of ~300–500 km s$^{-1}$. In Figure 3(a), two small EUV holes (circled) straddle the equator near $\phi \sim 120°$ and $\phi \sim 240°$; each has a counterpart open field region that is connected to the ecliptic plane (Figure 3(b)). The

## 2014 FEBRUARY 26 - 28

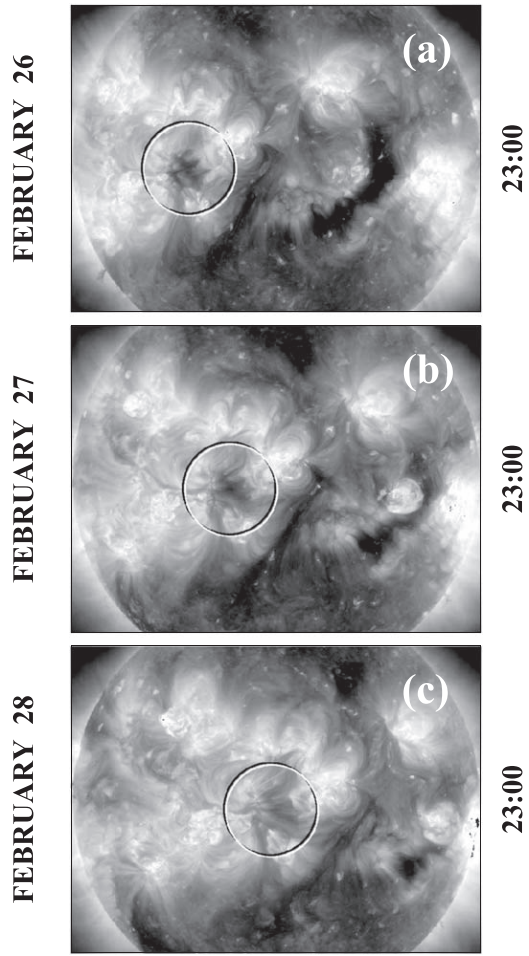

**FEBRUARY 26** — (a) 23:00

**FEBRUARY 27** — (b) 23:00

**FEBRUARY 28** — (c) 23:00

**Figure 2.** The Fe XIV 21.1 nm feature at $\phi \sim 130°$ in Figure 1(a) is tracked as it rotates toward central meridian during 2014 February 26–28. The sequence shows that the structure is a small coronal hole that becomes partially hidden beneath the loops rooted near its eastern edge.

PFSS-derived holes are embedded in positive unipolar areas and are characterized by large expansion factors, with the Earth-directed flux tubes having $f_{max} > 18$ (blue diamonds). The underlying photospheric fields have strengths of ~20–30 G and are associated with decaying ARs. The in situ observations (Figure 3(c)) show wind speeds of ~380–430 km s$^{-1}$ within the positive/outward IMF sector associated with the equatorial hole at $\phi \sim 120°$. The coronal hole at $\phi \sim 240°$ gives rise to a well-defined stream with speeds in the range ~380–490 km s$^{-1}$, which is likewise embedded in a positive IMF sector. In both cases, the O$^{7+}$/O$^{6+}$ ratios remain relatively low (~0.05–0.15) across the interiors of the holes, while rising steeply near their edges; the Fe/O ratios vary from a "floor" of ~0.2 to values as high as ~0.4. Figure 3(d) shows that the slow wind from both equatorial holes is characterized by high levels of Alfvénic fluctuations, with $|C_{vB}|$ reaching values of ~0.7.

### 3.3. CR 2150: 2014 May

Figure 4 shows the distribution of Fe XIV emission, the WSO photospheric field with open field areas overplotted, and the variation of the near-Earth solar wind during CR 2150.

## WSO 2148 (2014 MARCH 11 – APRIL 7)

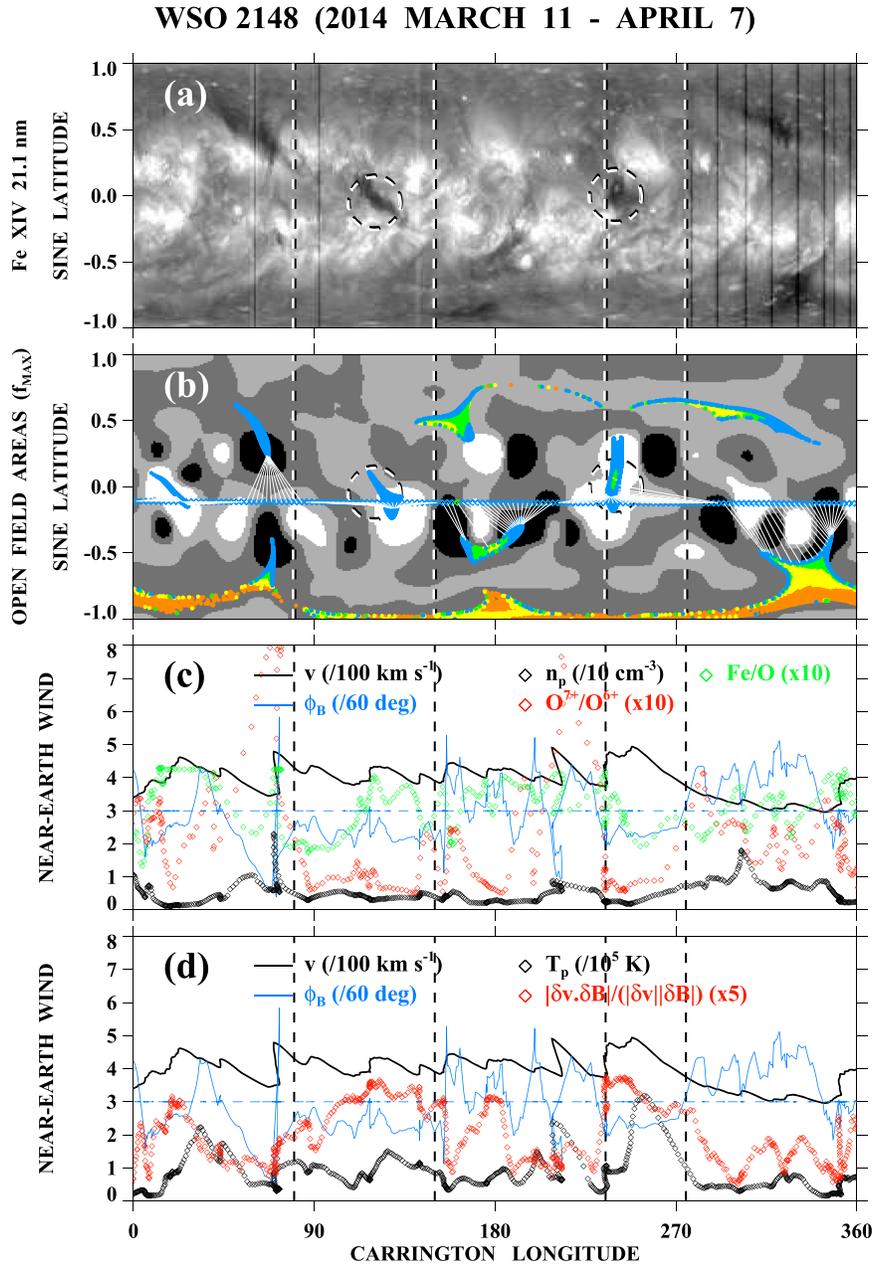

**Figure 3.** Two equatorial holes that gave rise to wind speeds in the range ~380–490 km s$^{-1}$ during CR 2148. (a) Distribution of 21.1 nm emission. The circled holes have equatorial widths of $\Delta\phi \sim 7°$. (b) WSO photospheric field with PFSS-derived open field regions overplotted (color-coding as in Figure 1(b)). (c) Near-Earth $v$, $n_p$, O$^{7+}$/O$^{6+}$, Fe/O, and IMF longitude angle, plotted in time-reversed order. (d) Variation of $T_p$ and the Alfvénicity parameter $|C_{vB}| \equiv |\delta v \cdot \delta B|/(|\delta v||\delta B|)$. The relatively slow wind from the two equatorial holes exhibits high Alfvénicity, with $|C_{vB}|$ reaching values of ~0.7.

Comparing Figures 3(a) and 4(a), we see that the earlier equatorial hole at $\phi \sim 240°$ has evolved into a long, thin structure that extends from the north polar region well into the southern hemisphere. The corresponding open field region is again characterized by very large expansion factors ($f_{max} > 18$: blue pixels) and is connected to the ecliptic plane over the longitude range ~215°–290° (Figure 4(b)). The solar wind measurements show very low speeds ($v \sim 330$–340 km s$^{-1}$) and correspondingly low proton temperatures ($T_p \sim 0.05$ MK) at these longitudes. That this wind comes from the transequatorial hole is supported by the low values of O$^{7+}$/O$^{6+}$ (~0.05–0.1) and Fe/O (~0.2) across the main body of the stream (Figure 4(c)) and by the high overall level ($C_{vB} \sim 0.6$) of the Alfvénicity parameter (Figure 4(d)).

### 3.4. CR 2152: 2014 June–July

Figure 5 shows the situation two rotations later. An equatorial hole is again present within the positive unipolar area around $\phi \sim 260°$, where the photospheric field is as strong as ~40 G (Figures 5(a) and (b)). The narrow midlatitude corridor seen during CR 2150 is no longer visible in the 21.1 nm map, although it remains present in the PFSS extrapolation of the low-resolution WSO measurements (Figure 5(b)). The hole continues to be the source of a low-speed stream with $v \sim 300$–390 km s$^{-1}$, O$^{7+}$/O$^{6+}$ $\gtrsim 0.06$, Fe/O $\gtrsim 0.17$, and Alfvénicities reaching $|C_{vB}| \sim 0.6$ (Figures 5(c)–(d)). Again referring to the Fe XIV map, a much narrower equatorial hole may be seen near $\phi \sim 45°$, which is apparently the source of the very slow wind within the

## WSO 2150 (2014 MAY 4 - MAY 31)

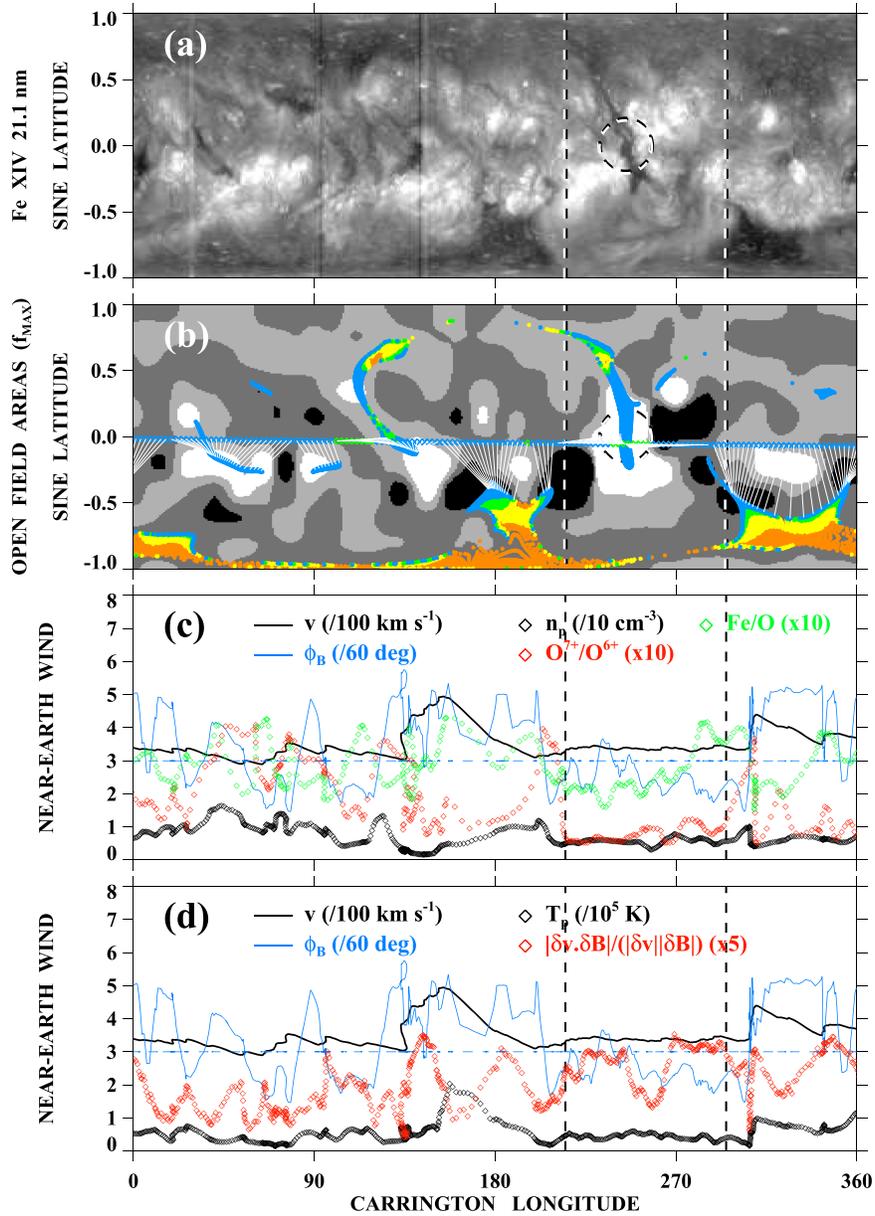

**Figure 4.** The equatorial hole at $\phi \sim 240°$ in Figure 3 has evolved into a long, thin structure connected to the north polar region during CR 2150. (a) Distribution of 21.1 nm emission. (b) WSO photospheric field with open field regions overplotted. (c) Near-Earth $v$, $\Phi_B$, $n_p$, $O^{7+}/O^{6+}$, and Fe/O. (d) Near-Earth proton temperature $T_p$ and Alfvénicity proxy $|C_{vB}|$. The low wind speeds associated with the narrow transequatorial hole are consistent with the large expansion factors ($f_{max} > 18$: blue diamonds in (b)) of the corresponding open field region. The coronal hole origin of this slow wind is also supported by the low $O^{7+}/O^{6+}$ ratios ($\sim 0.05–0.1$) and by the predominantly high levels of Alfvénicity. Throughout this CR, the variation of $T_p$ is well correlated with that of the wind speed (cc $\sim 0.65$).

positive IMF sector centered at longitude $\sim 60°$. The counterpart open field region in Figure 5(b) is connected to the ecliptic but does not extend as far northeastward as the observed coronal hole. The associated $O^{7+}/O^{6+}$ values are higher and more variable, and the Alfvénicities lower ($|C_{vB}| \lesssim 0.4$), than in the case of the wider equatorial hole at $\phi \sim 260°$. The variation of the proton temperature $T_p$ throughout this CR closely reflects that of the wind speed, with a correlation coefficient of cc $\simeq 0.70$.

### 3.5. CR 2154: 2014 August–September

After another two rotations (CR 2154), the equatorial hole near $\phi \sim 270°$ has increased noticeably in size (Figure 6(a));

as during CR 2150 (Figure 4(a)), a very narrow corridor connects it to high northern latitudes. Because of the effect of the photospheric differential rotation on the unipolar region in which the open flux is embedded (see Figure 6(b)), the corridor is now much more strongly inclined in the east–west direction than earlier. We remark that the tendency for polar hole extensions to rotate quasirigidly holds only if there are no large-scale polarity inversion lines nearby, as is more often the case during the declining phase of the cycle than near sunspot maximum (see Wang & Sheeley 1993). Here, the positive-polarity corridor is surrounded by strong negative-polarity regions and thus becomes increasingly

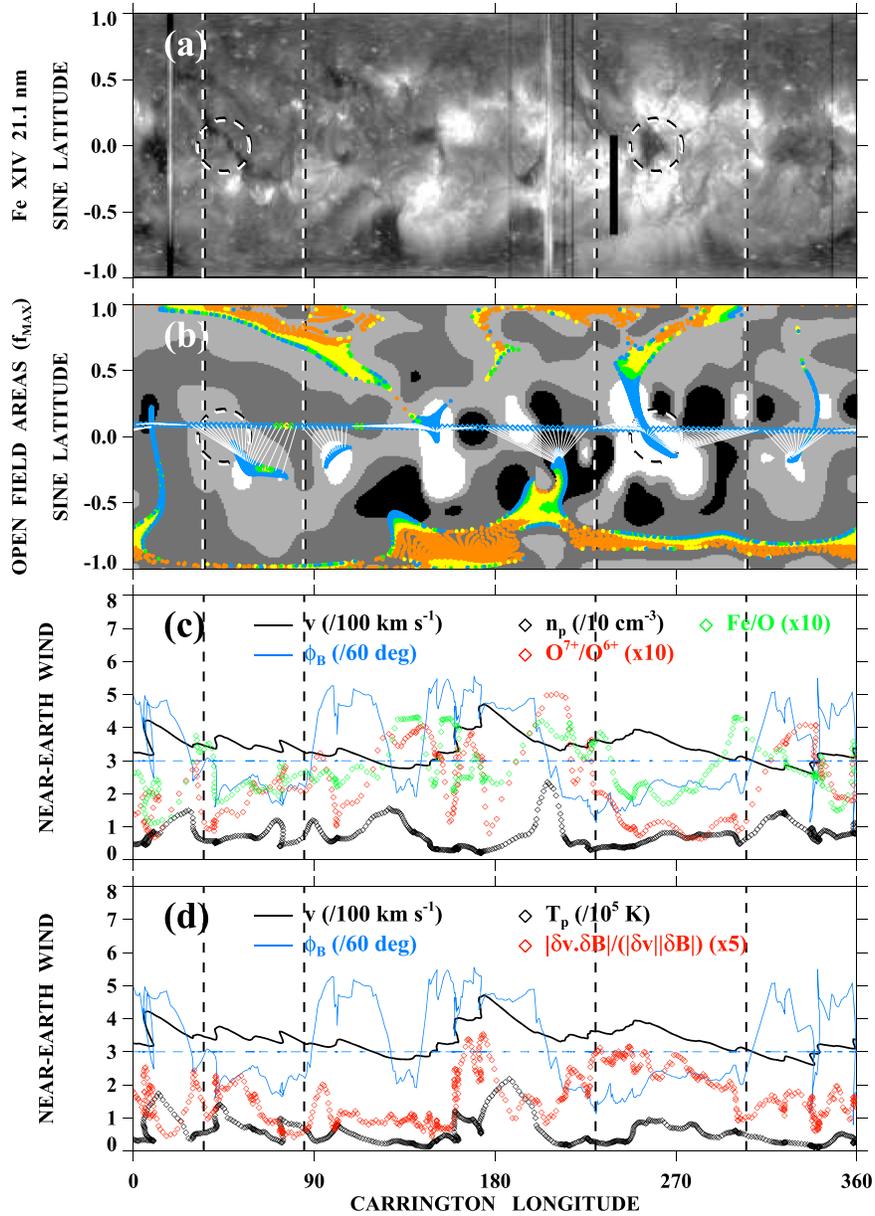

**Figure 5.** Two equatorial holes that were sources of slow wind during CR 2152. The hole of Figure 4 has widened to $\Delta\phi \sim 10°$ at the equator and gives rise to wind speeds $v \sim 300$–$390$ km s$^{-1}$ at longitudes $\sim 230°$–$300°$, with O$^{7+}$/O$^{6+} \sim 0.1$, Fe/O $\sim 0.2$, and $|C_{vB}| \sim 0.6$. The much narrower ($\Delta\phi \sim 3°$) equatorial hole near $\phi \sim 45°$ is the source of slow wind with $v \sim 330$–$370$ km s$^{-1}$, O$^{7+}$/O$^{6+}$ varying from $\sim 0.1$ to $\sim 0.2$, and Fe/O $\sim 0.2$; the Alfvénicity parameter has peak values $|C_{vB}| \sim 0.4$. Again, the variation of the proton temperature follows that of the wind speed.

slanted as the neighboring neutral lines are progressively sheared.

Consistent with the greater areal size of the equatorial hole during CR 2154, the associated wind stream shows higher speeds ($\sim 390$–$440$ km s$^{-1}$) than during CR 2152. Within the stream, the oxygen charge-state ratios are in the range $\sim 0.05$–$0.1$, while Fe/O fluctuates above a minimum value of $\sim 0.2$ (Figure 6(c)). As indicated by Figure 6(d), the associated Alfvénicity plateaus at a level as high as $|C_{vB}| \sim 0.7$.

This long-lived equatorial hole remains a persistent source of low-speed wind until early 2015. Figure 3 in Wang & Panasenco (2019) shows the hole and its associated wind stream during CR 2155 (2014 September–October).

### 3.6. CR 2164: 2015 May–June

Figure 7(a) displays the distribution of Fe XIV emission during CR 2164. A low-latitude extension of the embryo north polar hole crosses the equator at $\phi \sim 320°$ and then bends southeastward. The associated wind stream has a peak speed of only $\sim 400$ km s$^{-1}$, with O$^{7+}$/O$^{6+} \gtrsim 0.05$ and Fe/O $\gtrsim 0.25$ (Figure 7(c)). The PFSS extrapolation (Figure 7(b)) predicts a corresponding open field region whose Earth-directed flux tubes have expansion factors $f_{max} > 9$ (green and blue diamonds). Near longitude $90°$, the EUV map shows another positive-polarity equatorial hole, which is not connected to the north polar region and which is somewhat wider than the one at $\phi \sim 320°$. In this case, the associated wind speeds are

## WSO 2154 (2014 AUGUST 21 – SEPTEMBER 17)

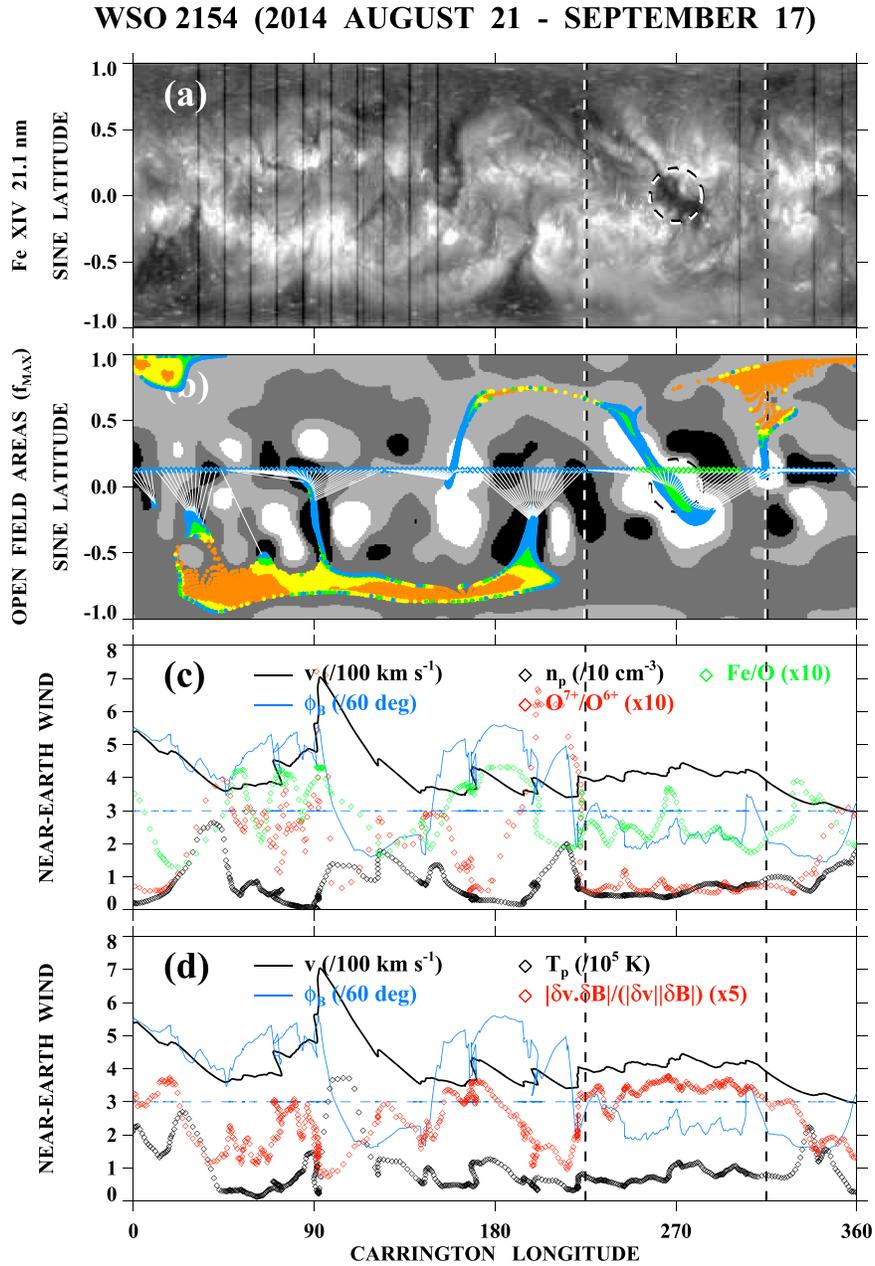

**Figure 6.** Distribution of 21.1 nm emission, WSO photospheric field with open field regions overplotted, and near-Earth solar wind measurements during CR 2154. The equatorial hole of Figure 5 has increased further in areal size while becoming attached to the north polar region by a narrow lane (as during CR 2150). The associated wind speeds are in the range ∼390–440 km s$^{-1}$, with O$^{7+}$/O$^{6+}$ ∼ 0.05–0.1 and Fe/O ≳ 0.2. The Alfvénicity level is as high as $|C_{vB}|$ ∼ 0.7.

considerably higher, reaching ∼580 km s$^{-1}$, a difference that might have been difficult to predict based solely on the morphological appearance of the two holes. However, the PFSS extrapolation (Figure 7(b)) indicates that the Earth-directed flux tubes from the open field region at $\phi \sim 90°$ have relatively low expansion factors ($4 < f_{max} < 9$: yellow diamonds), consistent with the higher speeds. This equatorial hole also has lower footpoint field strengths ($B_0 \sim 10$ G) than the one at $\phi \sim 320°$, which is embedded in photospheric fields as strong as $B_0 \sim 20$–40 G.

From Figure 7(d), we note that the slow wind from the transequatorial extension at $\phi \sim 320°$ has peak Alfvénicities $|C_{vB}| \sim 0.7$, comparable to those of the moderately fast wind from the equatorial hole at $\phi \sim 90°$ and of the high-speed stream from the lobe of the south polar hole at $\phi \sim 160°$.

However, the cores of both of the faster streams show significantly lower floor values of Fe/O (∼0.15 as compared to ∼0.25).

### 3.7. CR 2166: 2015 July–August

During 2015 June, a large northern-hemisphere AR (NOAA 12371) emerged to the east of the equatorward extension appearing in Figure 7(a). As may be seen from Figures 8(a) and (b), which display the distribution of 21.1 nm emission and the underlying photospheric field during CR 2166, the nearby injection of negative, leading-polarity flux shifted the equatorial hole westward and detached it from the north polar region. The narrow remnant hole is the probable source of the small wind stream near $\phi \sim 330°$, which has a maximum speed of ∼380 km s$^{-1}$ (Figure 8(c)), consistent with the very large expansion factors ($f_{max} > 18$: blue diamonds) characterizing the corresponding open field region (Figure 8(b)).

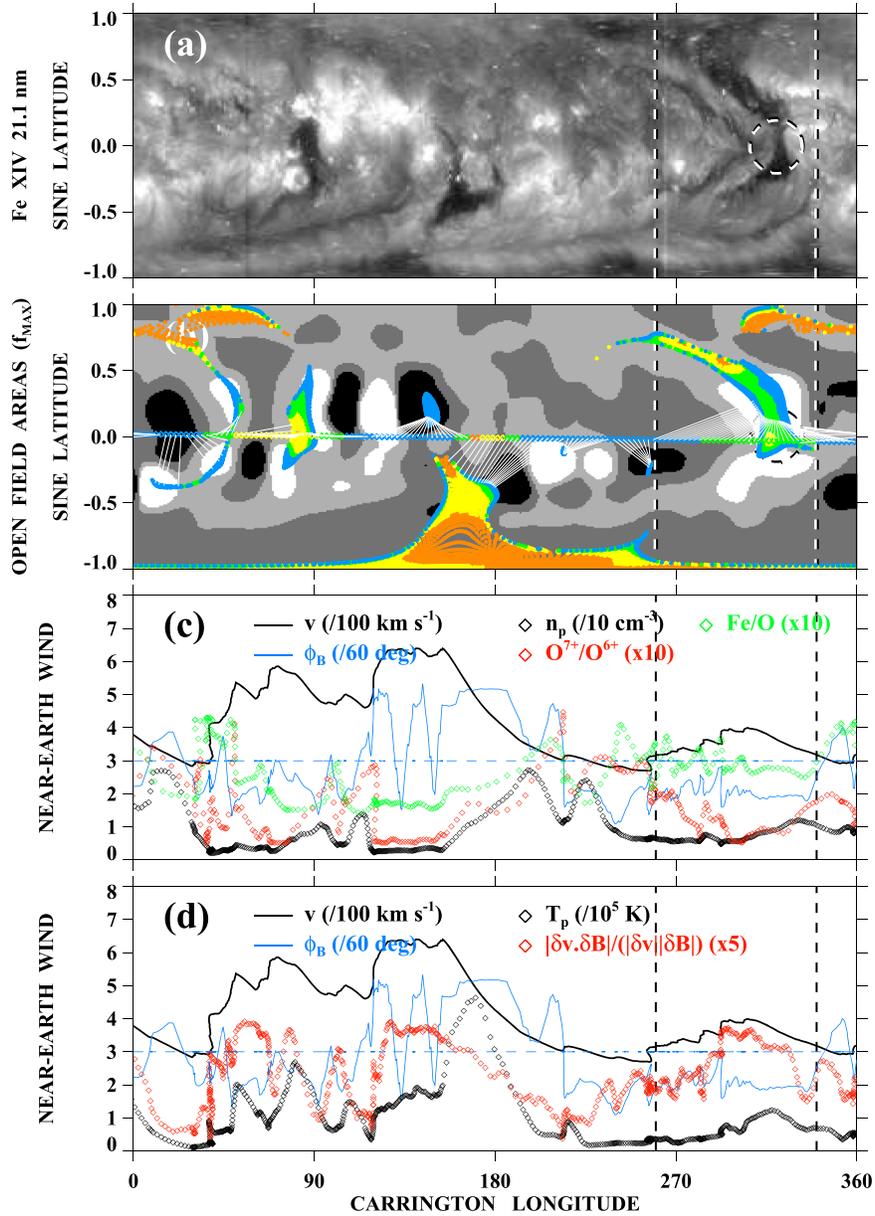

**Figure 7.** Distribution of 21.1 nm emission, WSO photospheric field with open field regions overplotted, and near-Earth solar wind measurements during CR 2164. The transequatorial extension of the north polar hole at $\phi \sim 320°$ is the source of a slow wind stream with $v \sim 320$–$400$ km s$^{-1}$, O$^{7+}$/O$^{6+} \gtrsim 0.05$, and Fe/O $\gtrsim 0.25$. Again, however, the Alfvénicity level reaches values as high as $|C_{vB}| \sim 0.7$, similar to the maximum values of the fast streams from the equatorial hole near $\phi \sim 90°$ and the south polar hole lobe near $\phi \sim 160°$. The lower speeds associated with the transequatorial extension are consistent with the larger expansion factors of its Earth-directed flux tubes ($f_{max} > 9$: green and blue diamonds).

The EUV map shows another equatorial hole at $\phi \sim 235°$, whose PFSS counterpart contains Earth-directed flux tubes with expansion factors $f_{max} > 9$. The corresponding in situ stream has a maximum speed of $\sim 480$ km s$^{-1}$.

The wind streams associated with the two equatorial holes are situated on opposite sides of a positive IMF sector extending over the longitude range $\sim 220°$–$360°$. According to Figure 8(b), the slow wind observed in the middle of this sector originates from two small spurs of the north polar hole. The succession of like-polarity streams are separated by local minima in $v$ accompanied by local maxima in O$^{7+}$/O$^{6+}$ and/or in Fe/O; these locations may represent pseudostreamer crossings (see Wang & Panasenco 2019).

The moderately slow wind from the equatorial hole at $\phi \sim 235°$ has high Alfvénicity ($|C_{vB}| \sim 0.6$), low O$^{7+}$/O$^{6+}$ ($\sim 0.05$), and low Fe/O ($\sim 0.2$); these values are similar to those characterizing the large, moderately fast stream from the transequatorial extension of the south polar hole at $\phi \sim 160°$. (It should be recalled, however, that the SWICS 2.0 measurements of O$^{7+}$/O$^{6+}$ have an instrumental floor of $\sim 0.05$, so that the actual minimum values of this parameter may be lower inside high-speed streams.) The slow wind from the narrow equatorial hole at $\phi \sim 330°$ has lower Alfvénicity ($|C_{vB}| \sim 0.4$), highly variable O$^{7+}$/O$^{6+}$, and Fe/O $\sim 0.2$.

## WSO 2166 (2015 JULY 14 - AUGUST 11)

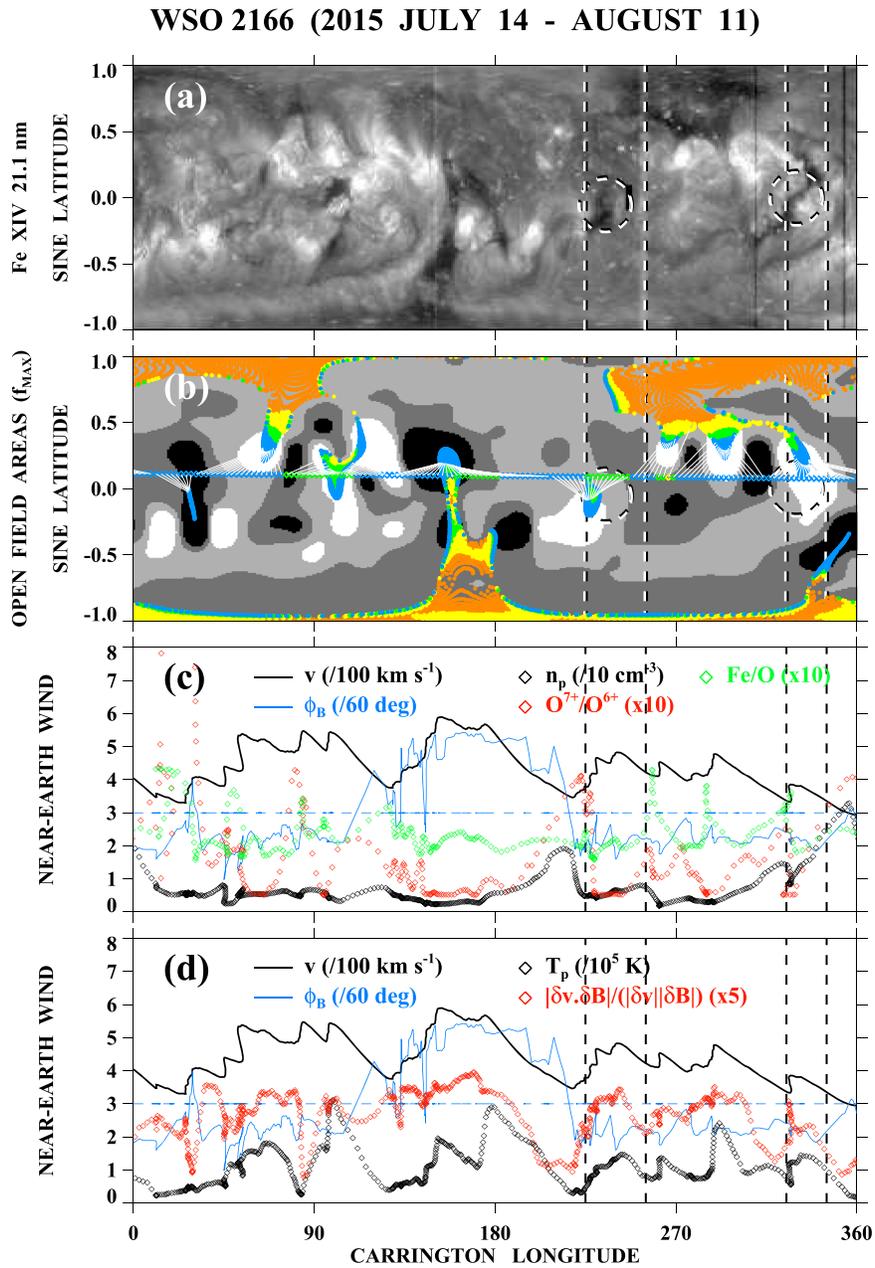

**Figure 8.** Distribution of 21.1 nm emission, WSO photospheric field with PFSS-derived open field regions overplotted, and in situ wind measurements for CR 2166. The low-speed (∼300–480 km s⁻¹) wind extending over longitudes ∼220°–360° consists of a series of streams, which originate from the equatorial holes at $\phi \sim 235°$ and $\phi \sim 330°$ and from two small spurs of the north polar hole. The streams are separated by pseudostreamers where $O^{7+}/O^{6+}$ and/or Fe/O show narrow peaks. The equatorial hole at $\phi \sim 235°$ has higher speeds and Alfvénicities, and lower values of $O^{7+}/O^{6+}$ and $f_{max}$, than the narrower hole at $\phi \sim 330°$.

### 3.8. CR 2174: 2016 February–March

Figure 9(a) shows the distribution of Fe XIV emission during CR 2174. A vertical extension of the north polar hole may be seen at $\phi \sim 245°$, which terminates near an AR complex in the southern hemisphere. According to the PFSS extrapolation in Figure 9(b), this narrow transequatorial hole, which has footpoint field strengths on the order of 15 G, is connected to the ecliptic over the longitude range ∼215°–275°. Figure 9(c) shows a clearly defined wind stream at this location, with a peak speed of ∼450 km s⁻¹, $O^{7+}/O^{6+} \sim 0.05$–0.15, and Fe/O ∼ 0.2–0.3. The velocities are consistent with the relatively large expansion factors predicted for the Earth-directed flux tubes ($f_{max} > 9$: green and blue diamonds in Figure 9(b)). As indicated by Figure 9(d), this wind stream has a high degree of Alfvénicity, with $|C_{vB}|$ reaching values of ∼0.7, exceeding even those associated with the faster wind stream from the neighboring equatorial hole at $\phi \sim 180°$.

### 3.9. CR 2178: 2016 June–July

The EUV map for CR 2178 (Figure 10(a)) shows a narrow (∆φ ∼ 5°), isolated equatorial hole at $\phi \sim 50°$. Its PFSS-derived counterpart in Figure 10(b) is embedded in a negative unipolar region and is connected to the ecliptic plane, with the Earth-directed flux tubes having footpoint field strengths of ∼10 G and expansion factors $f_{max} > 18$ (blue diamonds). The corresponding wind stream, which is confined within a narrow inward IMF sector, has a maximum speed of order 450 km s⁻¹, $O^{7+}/O^{6+} \sim 0.05$–0.1, and Fe/O ∼ 0.17–0.2 (Figure 10(c)). The Alfvénicity level $|C_{vB}|$ peaks at ∼0.7, similar to the values

## WSO 2174 (2016 FEBRUARY 18 – MARCH 16)

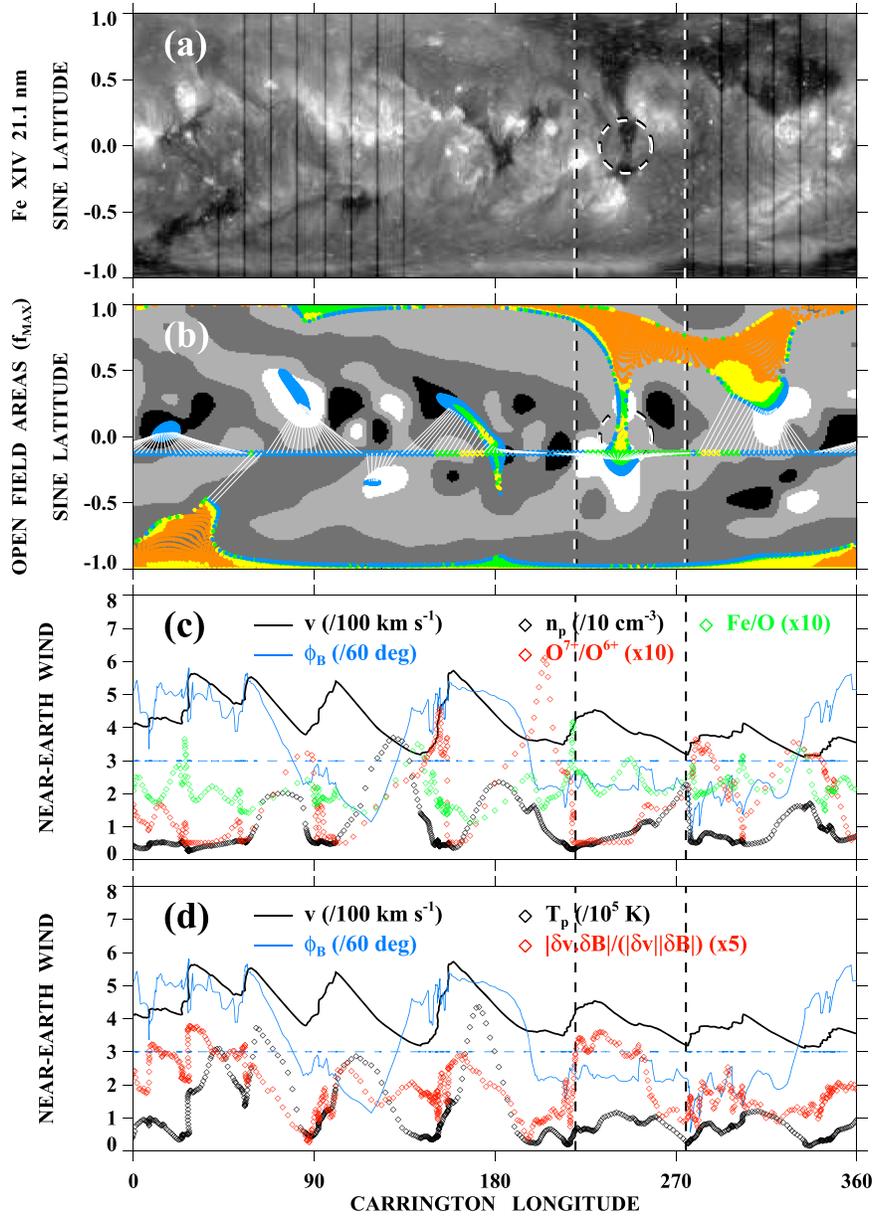

**Figure 9.** Synoptic map of Fe XIV intensity, open field regions derived from the WSO photospheric field, and near-Earth solar wind measurements during CR 2174. The EUV map shows a narrow ($\Delta\phi \sim 4°$) transequatorial extension of the north polar hole at $\phi \sim 245°$. The maximum wind speeds recorded during the passage of this hole are only of order 450 km s$^{-1}$, while O$^{7+}$/O$^{6+} \sim 0.05$–$0.15$ and Fe/O $\sim 0.2$–$0.3$. The Alfvénicity of this wind stream attains peak values of $|C_{vB}| \sim 0.7$, even higher than those of the moderately fast stream from the neighboring equatorial hole at $\phi \sim 180°$. The higher speeds associated with the latter hole are consistent with the smaller expansion factors of its Earth-directed flux tubes (yellow diamonds in (b)).

attained in the faster streams centered near $\phi \sim 150°$ and $\phi \sim 250°$ (Figure 10(d)).

### 3.10. CR 2195: 2017 September–October

Figure 11(a) displays the distribution of 21.1 nm emission during CR 2195. Around $\phi \sim 115°$, a coronal hole has formed in the gap between the ARs on opposite sides of the equator. From Figure 11(b), we see that the hole is embedded inside a negative-polarity area, with the counterpart open field region containing rapidly diverging flux connected to the ecliptic plane. The associated wind stream(s) have low speeds ($\sim$400 km s$^{-1}$), O$^{7+}$/O$^{6+}$ ratios of $\sim$0.05–0.1, and Fe/O $\gtrsim$ 0.2 (Figure 11(c)). As in our previous examples, this slow wind has high Alfvénicity

($|C_{vB}| \sim 0.7$), matching the levels shown by the very fast streams near longitudes 180° and 360° (Figure 11(d)). However, even though the Alfvénicity levels are similar, the minimum values of Fe/O are significantly lower ($\sim$0.13–0.16) within the cores of the two high-speed streams. We also note that the footpoint field strengths associated with these very fast streams are only $\sim$3–6 G, whereas the photospheric fields underlying the low-speed wind at longitudes $\sim$75°–135° are as strong as $\sim$20–30 G. The variation of $T_p$ over the CR tracks that of $v$ (cc $\simeq$ 0.79).

### 4. Summary and Discussion

The prevailing paradigm is that the solar wind and its sources are bimodal, with fast wind coming from coronal holes and slow

## WSO 2178 (2016 JUNE 6 - JULY 3)

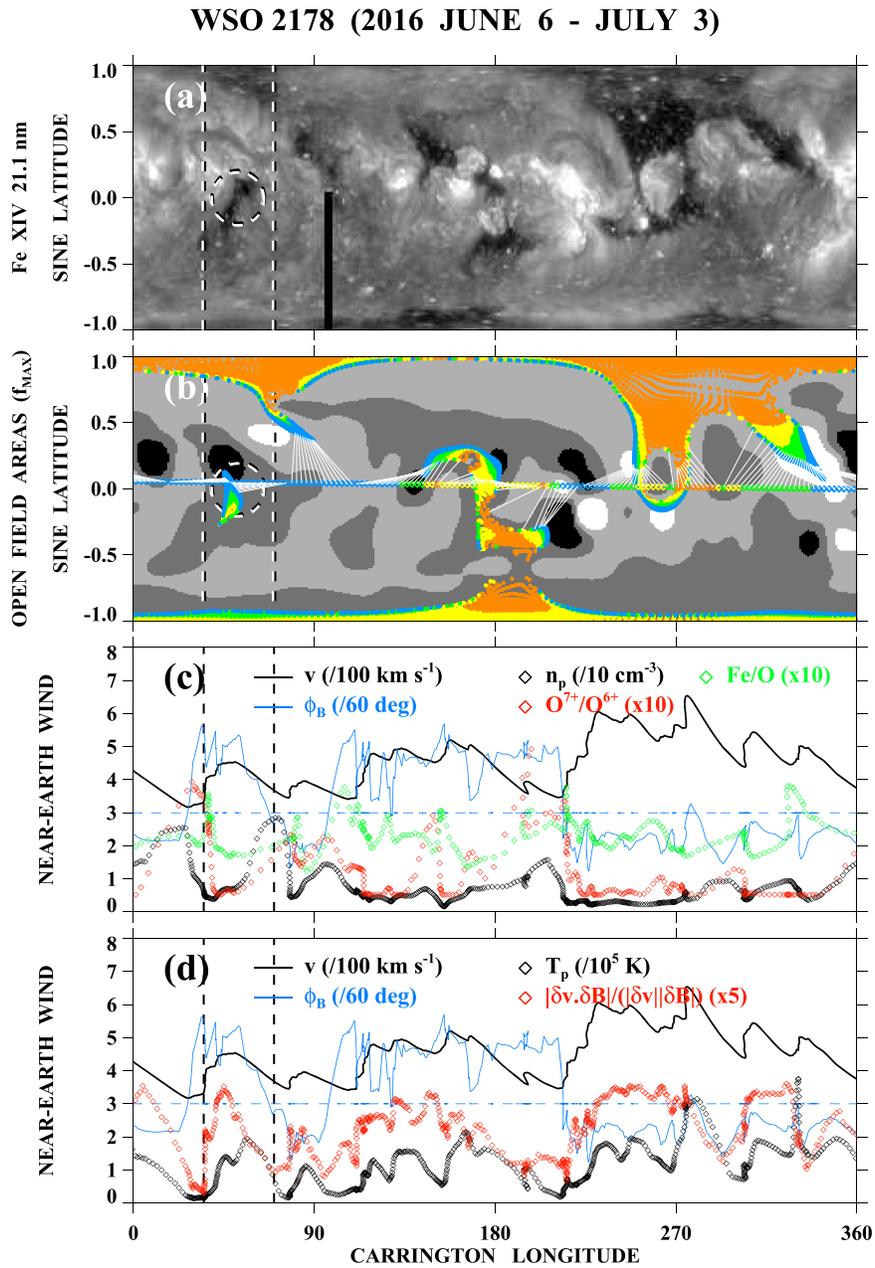

**Figure 10.** During CR 2178, an equatorial hole at longitude $\sim 50°$ gives rise to a small wind stream with maximum speeds of order 450 km s$^{-1}$, O$^{7+}$/O$^{6+}$ $\gtrsim$ 0.05, and Fe/O $\sim$ 0.2. The associated Alfvénicity rises to a peak value of $|C_{vB}| \sim 0.7$, matching the level shown by the faster streams near $\phi \sim 150°$ and $\phi \sim 250°$.

wind originating from streamers. To demonstrate that this picture is inconsistent with observations, we have identified examples of equatorial EUV holes during 2014–2017 that were almost certainly the source of low-speed ($\sim$300–450 km s$^{-1}$) wind at Earth. As in the case of most in-ecliptic wind streams, interactions during the transit to 1 au will have somewhat reduced the original peak speeds, while raising the floor speeds. Our focus was on equatorial coronal holes because they are likely to be magnetically connected to Earth, so that the dependence on the PFSS extrapolations is minimized; the equatorial location also increases the probability that Earth samples the interior of the hole and not just its boundaries. Strong independent support for our source identifications was provided by the Alfvénicity measurements.

The basic properties of our equatorial holes and their associated wind may be summarized as follows.

1. As observed in Fe XIV 21.1 nm images, the equatorial holes that were sources of low-speed wind tended to be relatively narrow, with longitudinal widths of $\Delta\phi \sim 3°$– 10° at the equator. They were located in and around low-latitude ARs and their remnants.

2. The O$^{7+}$/O$^{6+}$ ratios within the slow wind streams were generally in the range $\sim$0.05–0.15, consistent with a coronal hole origin. The values tended to increase rapidly toward the edges of the streams; near sector boundaries, O$^{7+}$/O$^{6+}$ rose to values typically on the order of 0.4.

3. The Fe/O ratios (a proxy for the FIP effect) were mostly in the range $\sim$0.2–0.3, with $\sim$0.2 being the typical

# GONG2 2195 (2017 SEPTEMBER 12 – OCTOBER 10)

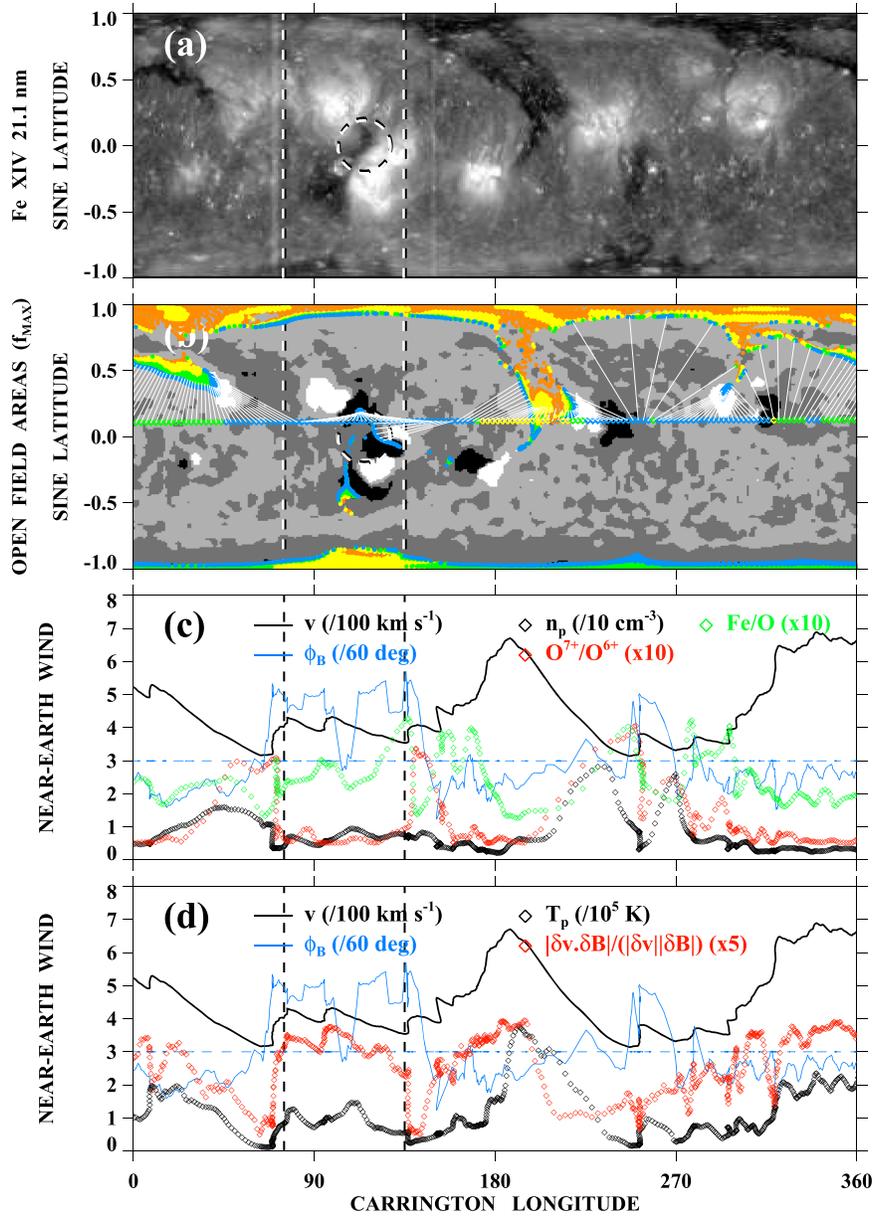

**Figure 11.** During CR 2195, an equatorial hole is seen forming near $\phi \sim 115°$ between ARs on opposite sides of the equator. The associated wind speeds are on the order of 400 km s$^{-1}$. In contrast, the transequatorial extension of the north polar hole (near $\phi \sim 180°$) gives rise to a high-speed stream with speeds peaking at ~670 km s$^{-1}$. In both cases, the Alfvénicity level reaches $|C_{vB}| \sim 0.7$; however, Fe/O falls to much lower values within the fast stream. As shown in (b), the Earth-directed flux tubes from the transequatorial extension have much smaller expansion factors ($f_{max} < 9$: yellow and orange diamonds) than those from the open field region at $\phi \sim 115°$ ($f_{max} > 18$: blue diamonds). The photospheric field measurements are from GONG2.

"floor" value within the low-speed streams. The minimum values were lower (Fe/O ∼ 0.15) in some high-speed streams.

4. The slow wind from the equatorial holes was characterized by large values of the Alfvénicity proxy $|C_{vB}| \equiv |\delta v \cdot \delta B|/(|\delta v||\delta B|)$ (∼0.6–0.7). These values were similar to those observed in high-speed streams (see Figures 7(d)–11(d)).

5. In general, $|C_{vB}|$ tended to fall steeply near the edges of both slow and fast wind streams. This behavior is consistent with the presence of a non-coronal hole (helmet streamer or pseudostreamer) source of slow wind, but we were unable to ascertain the exact location of the hole–streamer interface, whose structure may have been modified by interactions during the Sun–Earth transit.

6. During the 2014 sunspot maximum, the small equatorial holes that gave rise to slow wind had footpoint field strengths typically on the order of 30 G. Although the footpoint fields in such holes tended to weaken as sunspot activity declined, they remained stronger than the fields of only a few gauss associated with large high-speed streams.

7. As expected, the proton temperature was generally well correlated with $v$ (cc ∼ 0.7), being lower inside the slow wind originating from small equatorial holes than inside faster wind streams. The smallest values of $T_p$ occurred

## GONG2 PHOTOSPHERIC FIELD & OPEN FLUX (CR 2047 - 2213)

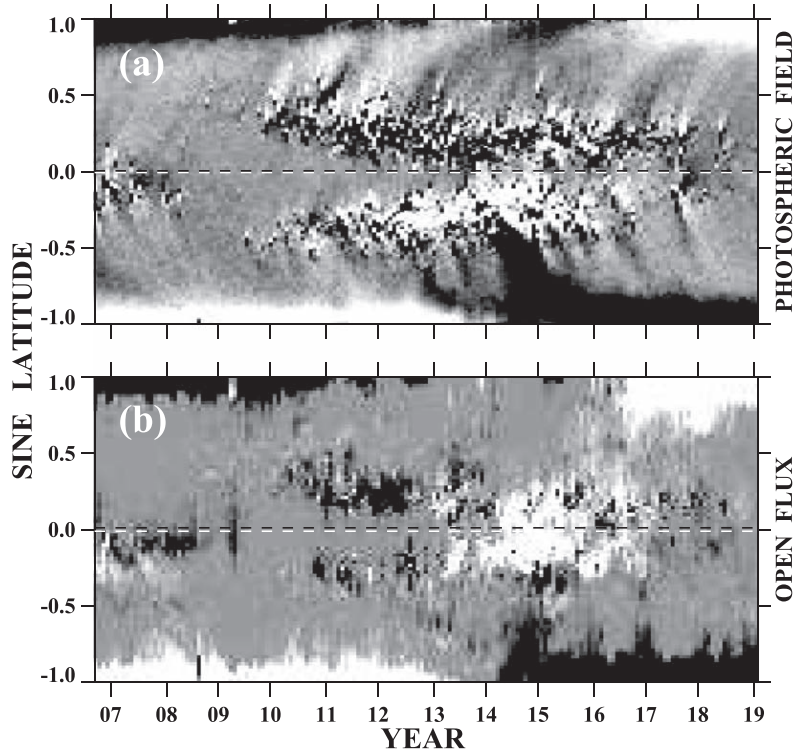

**Figure 12.** Latitude–time plots of (a) the photospheric field and (b) the open flux at the solar surface, derived from GONG2 maps for CR 2047–2213. Gray scale for the longitudinally averaged photospheric field ranges from $B_r < -3$ G (black) to $B_r > +3$ G (white); the GONG2 fields have been scaled upward by 1.65 to make them consistent with measurements from NSO/SOLIS. The longitudinally averaged distribution of open flux was determined by applying a PFSS extrapolation to the photospheric field maps; gray scale ranges from $-0.5$ G (black) to $+0.5$ G (white). From (b), it is apparent why most of our equatorial coronal holes were from the period 2014–2015 and why the great majority had positive polarity. The concentration of positive-polarity open flux around the equator at this time reflects the concentration of sunspot activity in the southern hemisphere, where the leading/equatorward polarity for cycle 24 was positive.

toward the edges of coronal holes and their associated streams, including near sector boundaries and stream interfaces.

8. Although not plotted here, the level of velocity fluctuations $\delta v$, or standard deviation of $v$ over timescales of $\sim 1$ hr, tended to be well correlated with $T_p$ (cc $\sim 0.7$), as found by Ko et al. (2018).

9. The Earth-directed flux tubes from the equatorial holes associated with slow wind had large expansion factors ($f_{max} > 9$). In contrast, equatorial holes that gave rise to fast wind at Earth (such as those at $\phi \sim 90°$ in Figure 7, $\phi \sim 180°$ in Figure 9, $\phi \sim 270°$ in Figure 10, and $\phi \sim 180°$ in Figure 11) had relatively small expansion factors ($f_{max} < 9$).

Our conclusions are consistent with the recent study of D'Amicis et al. (2019), who identified episodes of high Alfvénicity in slow wind at 1 au during the 2000–2002 sunspot maximum. They attributed this wind to small low-latitude coronal holes, and suggested that the low speeds were associated with the effect of superradial expansion (see Wang 1994, 2017; Wang & Sheeley 2003).

Small equatorial coronal holes are most frequently encountered during the maximum and early post-maximum phases of the cycle, when sunspot activity is high and centered at low latitudes. Our results and those of D'Amicis et al. (2019) confirm that such coronal holes are a major contributor to the slow wind in the ecliptic during this period. As the rate of AR emergence declines, the unipolar regions formed from their remnants and the embedded low-latitude holes tend to increase in areal size, leading to higher average wind speeds at Earth.

That 11 of the 14 equatorial holes analyzed here were embedded in positive-polarity regions is not a coincidence, but can be attributed to the fact that sunspot activity was heavily concentrated in the southern hemisphere during the maximum of cycle 24. As a result, the equatorial zone was dominated by positive-polarity flux originating from the leading/equatorward sectors of southern-hemisphere ARs. Figure 12 displays the latitude–time distribution of the GONG2 photospheric field from 2006 to the present, together with the corresponding distribution of open flux derived by applying a PFSS extrapolation to the GONG2 measurements and averaging over longitude. During 2014–2015, the positive-polarity open flux is seen to be densely clustered around the equator (Figure 12(b)); in contrast, the negative-polarity open flux is concentrated toward the south pole, reflecting the giant "poleward surge" of trailing-polarity AR flux occurring in that hemisphere (Figure 12(a)). The relative amount of negative-polarity open flux near the equator increases from 2016 onward, when the main locus of sunspot activity shifts to the northern hemisphere.

Based on our results for equatorial holes, it is natural to assume that coronal holes at higher latitudes may also be sources of slow solar wind, both in and out of the ecliptic plane. According to the PFSS extrapolations, a major contribution to the slow wind near Earth around solar minimum comes from just inside the polar hole boundaries. As demonstrated by ACE/SWICS 1.1 measurements during 2007–2009, when the median value of $O^{7+}/O^{6+}$ fell to only $\sim 0.06$ despite the fact that the HCS lay near the ecliptic (Wang 2016), ion charge-state composition is not necessarily a reliable discriminator between coronal hole and streamer wind. By flying closer to the Sun, the *Parker Solar Probe* and *Solar Orbiter* missions may allow us to better resolve the boundaries between the different types of wind. Furthermore, we anticipate that these new observations may help to confirm that the bulk of the "young" solar wind, both slow and fast, is characterized by high levels of Alfvénic fluctuations, as found by Stansby et al. (2019) using *Helios* data from four solar cycles ago.

We are indebted to L. Hutting for constructing the AIA synoptic maps displayed in this paper. This work was supported by the NASA H-GI Open and NASA/NOAA/NSF HSWO2R programs and by the Chief of Naval Research.

**ORCID iDs**

Y.-M. Wang  https://orcid.org/0000-0002-3527-5958
Y.-K. Ko  https://orcid.org/0000-0002-8747-4772

**References**

Abbo, L., Ofman, L., Antiochos, S. K., et al. 2016, SSRv, 201, 55
Antiochos, S. K., Mikić, Z., Titov, V. S., Lionello, R., & Linker, J. A. 2011, ApJ, 731, 112
Arge, C. N., & Pizzo, V. J. 2000, JGR, 105, 10465
Cohen, O. 2015, SoPh, 290, 2245
Cranmer, S. R., van Ballegooijen, A. A., & Edgar, R. J. 2007, ApJS, 171, 520
D'Amicis, R., & Bruno, R. 2015, ApJ, 805, 84
D'Amicis, R., Matteini, L., & Bruno, R. 2019, MNRAS, 483, 4665
Ko, Y.-K., Roberts, D. A., & Lepri, S. T. 2018, ApJ, 864, 139
Marsch, E., Mühlhäuser, K.-H., Rosenbauer, H., Schwenn, R., & Denskat, K. U. 1981, JGR, 86, 9199
Neugebauer, M., Forsyth, R. J., Galvin, A. B., et al. 1998, JGR, 103, 14587
Oran, R., Landi, E., van der Holst, B., et al. 2015, ApJ, 806, 55
Poduval, B. 2016, ApJL, 827, L6
Riley, P., Linker, J. A., Mikić, Z., et al. 2006, ApJ, 653, 1510
Roberts, D. A., Goldstein, M. L., Klein, L. W., & Matthaeus, W. H. 1987a, JGR, 92, 12023
Roberts, D. A., Klein, L. W., Goldstein, M. L., & Matthaeus, W. H. 1987b, JGR, 92, 11021
Sheeley, N. R., Jr., Lee, D. D.-H., Casto, K. P., Wang, Y.-M., & Rich, N. B. 2009, ApJ, 694, 1471
Stakhiv, M., Landi, E., Lepri, S. T., Oran, R., & Zurbuchen, T. H. 2015, ApJ, 801, 100
Stansby, D., Horbury, T. S., & Matteini, L. 2019, MNRAS, 482, 1706
Wang, Y.-M. 1994, ApJL, 437, L67
Wang, Y.-M. 2016, ApJ, 833, 121
Wang, Y.-M. 2017, ApJ, 841, 94
Wang, Y.-M., Ko, Y.-K., & Grappin, R. 2009, ApJ, 691, 760
Wang, Y.-M., & Panasenco, O. 2019, ApJ, 872, 139
Wang, Y.-M., & Sheeley, N. R., Jr. 1990, ApJ, 355, 726
Wang, Y.-M., & Sheeley, N. R., Jr. 1993, ApJ, 414, 916
Wang, Y.-M., & Sheeley, N. R., Jr. 1995, ApJL, 447, L143
Wang, Y.-M., & Sheeley, N. R., Jr. 2003, ApJ, 587, 818
Zhao, L., Landi, E., Zurbuchen, T. H., Fisk, L. A., & Lepri, S. T. 2014, ApJ, 793, 44
Zhao, L., Zurbuchen, T. H., & Fisk, L. A. 2009, GeoRL, 36, L14104